\documentclass[aps,prl,preprint,tightenlines,superscriptaddress,showpacs,byrevtex]{revtex4}
\usepackage{graphicx} 
\usepackage{dcolumn} 

\graphicspath{{ps}}

\begin{document}


\preprint{\vbox{ \hbox{   }
                 \vspace*{1.5cm}
                 \hbox{BELLE-CONF-0626}
}}

\title{ \quad\\[0.5cm]  Observation of  $B \to \phi \phi K$ Decays}

\affiliation{Budker Institute of Nuclear Physics, Novosibirsk}
\affiliation{Chiba University, Chiba}
\affiliation{Chonnam National University, Kwangju}
\affiliation{University of Cincinnati, Cincinnati, Ohio 45221}
\affiliation{University of Frankfurt, Frankfurt}
\affiliation{The Graduate University for Advanced Studies, Hayama} 
\affiliation{Gyeongsang National University, Chinju}
\affiliation{University of Hawaii, Honolulu, Hawaii 96822}
\affiliation{High Energy Accelerator Research Organization (KEK), Tsukuba}
\affiliation{Hiroshima Institute of Technology, Hiroshima}
\affiliation{University of Illinois at Urbana-Champaign, Urbana, Illinois 61801}
\affiliation{Institute of High Energy Physics, Chinese Academy of Sciences, Beijing}
\affiliation{Institute of High Energy Physics, Vienna}
\affiliation{Institute of High Energy Physics, Protvino}
\affiliation{Institute for Theoretical and Experimental Physics, Moscow}
\affiliation{J. Stefan Institute, Ljubljana}
\affiliation{Kanagawa University, Yokohama}
\affiliation{Korea University, Seoul}
\affiliation{Kyoto University, Kyoto}
\affiliation{Kyungpook National University, Taegu}
\affiliation{Swiss Federal Institute of Technology of Lausanne, EPFL, Lausanne}
\affiliation{University of Ljubljana, Ljubljana}
\affiliation{University of Maribor, Maribor}
\affiliation{University of Melbourne, Victoria}
\affiliation{Nagoya University, Nagoya}
\affiliation{Nara Women's University, Nara}
\affiliation{National Central University, Chung-li}
\affiliation{National United University, Miao Li}
\affiliation{Department of Physics, National Taiwan University, Taipei}
\affiliation{H. Niewodniczanski Institute of Nuclear Physics, Krakow}
\affiliation{Nippon Dental University, Niigata}
\affiliation{Niigata University, Niigata}
\affiliation{University of Nova Gorica, Nova Gorica}
\affiliation{Osaka City University, Osaka}
\affiliation{Osaka University, Osaka}
\affiliation{Panjab University, Chandigarh}
\affiliation{Peking University, Beijing}
\affiliation{University of Pittsburgh, Pittsburgh, Pennsylvania 15260}
\affiliation{Princeton University, Princeton, New Jersey 08544}
\affiliation{RIKEN BNL Research Center, Upton, New York 11973}
\affiliation{Saga University, Saga}
\affiliation{University of Science and Technology of China, Hefei}
\affiliation{Seoul National University, Seoul}
\affiliation{Shinshu University, Nagano}
\affiliation{Sungkyunkwan University, Suwon}
\affiliation{University of Sydney, Sydney NSW}
\affiliation{Tata Institute of Fundamental Research, Bombay}
\affiliation{Toho University, Funabashi}
\affiliation{Tohoku Gakuin University, Tagajo}
\affiliation{Tohoku University, Sendai}
\affiliation{Department of Physics, University of Tokyo, Tokyo}
\affiliation{Tokyo Institute of Technology, Tokyo}
\affiliation{Tokyo Metropolitan University, Tokyo}
\affiliation{Tokyo University of Agriculture and Technology, Tokyo}
\affiliation{Toyama National College of Maritime Technology, Toyama}
\affiliation{University of Tsukuba, Tsukuba}
\affiliation{Virginia Polytechnic Institute and State University, Blacksburg, Virginia 24061}
\affiliation{Yonsei University, Seoul}
  \author{K.~Abe}\affiliation{High Energy Accelerator Research Organization (KEK), Tsukuba} 
  \author{K.~Abe}\affiliation{Tohoku Gakuin University, Tagajo} 
  \author{I.~Adachi}\affiliation{High Energy Accelerator Research Organization (KEK), Tsukuba} 
  \author{H.~Aihara}\affiliation{Department of Physics, University of Tokyo, Tokyo} 
  \author{D.~Anipko}\affiliation{Budker Institute of Nuclear Physics, Novosibirsk} 
  \author{K.~Aoki}\affiliation{Nagoya University, Nagoya} 
  \author{T.~Arakawa}\affiliation{Niigata University, Niigata} 
  \author{K.~Arinstein}\affiliation{Budker Institute of Nuclear Physics, Novosibirsk} 
  \author{Y.~Asano}\affiliation{University of Tsukuba, Tsukuba} 
  \author{T.~Aso}\affiliation{Toyama National College of Maritime Technology, Toyama} 
% ...
\collaboration{The Belle Collaboration}


\noaffiliation

\begin{abstract}
We report the observation of the decay $B^{\pm} \to \phi \phi K^{\pm}$ and find evidence for $B^{0} \to \phi \phi K^{0}$.
These results are based on a 414 fb$^{-1}$ data sample that contains $449 \times 10^6$ $B\overline{B}$ pairs, collected with the Belle detector at the KEKB asymmetric-energy $e^+e^-$ (3.5 on 8 GeV) collider operating at the $\Upsilon(4S)$ resonance.
This is the first observation of a $b \to s \overline{s} s \overline{s} s$ transition.
The branching fractions for these decay modes are measured to be $\mathcal{B}(B^{\pm} \to \phi \phi K^{\pm}) = (3.2^{+0.6}_{-0.5} \pm 0.3) \times 10^{-6}$ and $\mathcal{B}(B^{0} \to \phi \phi K^{0}) = (2.3^{+1.0}_{-0.7} \pm 0.2) \times 10^{-6}$ for $\phi \phi$ invariant mass below 2.85 GeV/$c^2$.
The corresponding partial rate asymmetry for the charged $B$ mode is measured to be $\mathcal{A}_{CP}(B^\pm\to \phi \phi K^\pm) = 0.01^{+ 0.19}_{-0.16}\pm 0.02$.
Results for other related charmonium decay modes are also reported.
We also search for $CP$ asymmetry using the $\phi\phi$ candidates within the $\eta_c$ mass region.
The value obtained is $\mathcal{A}_{CP}(B^\pm \to \phi \phi K^\pm, M_{\phi \phi} \approx M_{\eta_c})= 0.15^{+0.16}_{-0.17} \pm 0.02$, which is consistent with no asymmetry.
\end{abstract}

\pacs{13.25.Ft, 13.25.Hw, 14.40.Nd}

\maketitle

\tighten

{\renewcommand{\thefootnote}{\fnsymbol{footnote}}}
\setcounter{footnote}{0}

Evidence of charmless $B \to \phi \phi K$ decays has been reported by the Belle collaboration using $85 \times 10^6$ $B\overline{B}$ pairs~\cite{Huang}.
In the standard model (SM), this decay channel requires the creation of an additional final $s \overline{s}$ quark pair in a $b \to s \overline{s} s$ process, such as $B \to \phi K$.
Although the statistical error of the previous $B \to \phi \phi K$ measurement was large, the branching fraction was found to be around 1/3 of that of $B \to \phi K$, which provided useful information for understanding quark fragmentation in $B$ decay.
Moreover, it indicated that with sufficient statistics the decay $B \to \phi \phi K$ could be used to search for a possible non-SM $CP$-violating phase in the $b \to s$ transition~\cite{Hazumi}.
Direct $CP$ violation could be enhanced as high as the 40\% level if there is sizable interference between the transitions due to non-SM physics and the decays via the $\eta_{c}$ resonance.\par

In this paper, we report the observation of charmless and charmful $B \to \phi \phi K$ decays based on a $414~{\rm fb}^{-1}$ data sample that contains 449 $\times 10^6 B\overline{B}$ pairs, collected with the Belle detector at the KEKB asymmetric-energy $e^+e^-$ (3.5 on 8~GeV) collider~\cite{KEKB} operating at the $\Upsilon(4S)$ resonance.\par

The Belle detector is a large-solid-angle magnetic spectrometer that consists of a silicon vertex detector (SVD), a 50-layer central drift chamber (CDC), 
an array of aerogel threshold \v{C}erenkov counters (ACC), a barrel-like arrangement of time-of-flight scintillation counters (TOF), and an electromagnetic calorimeter comprised of CsI(Tl) crystals (ECL) located inside a super-conducting solenoid coil that provides a 1.5~T magnetic field.
An iron flux-return located outside of the coil is instrumented to detect $K_L^0$ mesons and to identify muons (KLM).
The detector is described in detail elsewhere~\cite{Belle}.
Two inner detector configurations were used.
A 2.0 cm radius beampipe and a 3-layer silicon vertex detector (SVD1) were used for the first sample of 152 $\times 10^6 B\overline{B}$ pairs, while a 1.5 cm radius beampipe, a 4-layer silicon detector (SVD2) and a small-cell inner drift chamber were used to record the remaining 297 $\times 10^6 B\overline{B}$ pairs~\cite{Ushiroda}.\par

Charged kaons are required to have impact parameters within $\pm$2 cm of the interaction point (IP) along the $z$-axis (anti-parallel to the positron direction) and within 0.2 cm in the transverse plane.
Each track is identified as a kaon or a pion according to a $K/\pi$ likelihood ratio, $\mathcal{R}(K/\pi) = \mathcal{L}_K/(\mathcal{L}_K+\mathcal{L}_\pi)$, where $\mathcal{L}_K/\mathcal{L}_\pi$ is the likelihood of kaons/pions derived from the responses of TOF and ACC systems and the energy loss measurements from the CDC.
The likelihood ratio is required to exceed 0.6 for kaon candidates; within the momentum range of interest, this requirement is 88\% efficient for kaons and has a misidentification rate for pions of 8.5\%.
Neutral kaons are reconstructed via the decay $K^{0}_{S} \to \pi^{+} \pi^{-}$ and have an invariant mass 0.482 GeV/$c^2 < M_{\pi^{+} \pi^{-}} <$ 0.514 GeV/$c^2$ ($\pm4 \sigma$ mass resolution).
The $\pi^{+} \pi^{-}$ vertex is required to be displaced from the IP and the flight direction must be consistent with a $K^{0}_{S}$ that originated from the IP.
The required displacement increases with the momentum of the $K^0_S$ candidate.\par

$B$ meson candidates are reconstructed in the five kaon final state.
Two kinematic variables are used to distinguish signal candidates from backgrounds: the beam-energy constrained mass $M_{\rm bc} = \sqrt{E^{2}_{\rm beam} - |\vec{P}_{\rm recon}|^{2}}$ and the energy difference $\Delta E = E_{\rm recon} - E_{\rm beam}$, where $E_{\rm beam}$ is the beam energy, and $E_{\rm recon}$ and $\vec{P}_{\rm recon}$ are the reconstructed energy and momentum of the signal candidate in the $\Upsilon(4S)$ rest frame. The resolution of $M_{\rm bc}$ is approximately 2.8 MeV/$c^2$, dominated by the beam energy spread, while the $\Delta E$ resolution is around 10 MeV.
Candidates with five kaons within the region $|\Delta E| <$ 0.2 GeV and 5.2 GeV/$c^2 < M_{\rm bc}$ are selected for further consideration.
The signal region is defined as 5.27 GeV/$c^2 < M_{\rm bc} <$ 5.29 GeV/$c^2$ and $|\Delta E| <$ 0.05 GeV.\par

The dominant backgrounds are $e^{+}e^{-} \to q\overline{q}$ ($q=u, d, s, c$) continuum events.
Event topology and $B$ flavor tagging are used to distinguish the jet-like continuum events and the spherically distributed $B\overline{B}$ events. 
Seven event-shape variables are combined into a single Fisher discriminant~\cite{fisher}.
The Fisher variables include the angle between the thrust axis of the $B$ candidate and the thrust axis of the rest of the event ($\cos\theta_{T}$),  five modified Fox-Wolfram moments~\cite{SFW}, and a measure of the momentum transverse to the event thrust axis ($S_\perp$)~\cite{spher}.
Two other variables that are uncorrelated with the Fisher discriminant and help to distinguish signal from the continuum are $\cos\theta_{B}$, where $\theta_{B}$ is the angle between the $B$ flight direction and the beam direction in the $e^+e^-$ center of mass frame, and $\Delta z$, the $z$ vertex difference between the signal $B$ candidate and its accompanying $B$.
We form signal and background probability density functions (PDFs) for the Fisher discriminant, $\cos\theta_{B}$ and $\Delta z$ using the signal Monte Carlo (MC) events and sideband data (5.2 GeV/$c^2 < M_{\rm bc} <$ 5.26 GeV/$c^2$), respectively. 
The products of the PDFs for these variables give signal and background likelihoods $\mathcal{L}_{S}$ and $\mathcal{L}_{BG}$ for each candidate, allowing a selection to be applied to the likelihood ratio $\mathcal{R}=\mathcal{L}_{S}/(\mathcal{L}_{S}+\mathcal{L}_{BG})$.\par

Additional background discrimination is provided by the quality of the $B$ flavor tagging of the  accompanying $B$ meson.
The standard Belle flavor tagging package~\cite{TaggingNIM} gives two outputs: a discrete variable indicating the flavor of the tagging $B$ and dilution factor $r$, which ranges from zero for no flavor information to unity for unambiguous flavor assignment.
The continuum background is reduced by applying a selection requirement of $\mathcal{R}$ for events in each $r$ region according to the figure of merit defined as $N_{S}/\sqrt{N_{S} + N_{BG}}$, where $N_{S}$ denotes the expected $\phi \phi K$ signal yields based on MC simulation and the branching fraction reported in our previous measurements, and $N_{BG}$ denotes the expected $q\overline{q}$ yields from sideband data.
This requirement removes (61-81)\% of the continuum background while retaining (80-92)\% of the signal, and depends on the decay channel ($\phi \phi K$ or $\phi \phi K^0$) and the SVD configuration during the measurement (SVD1 or SVD2).
Backgrounds from other $B$ decays are investigated using a large MC sample and are found to be negligible after the $\mathcal{R}$ requirement.\par

The signal yields are extracted by applying an unbinned extended maximum likelihood (ML) fit to the events with $M_{\rm bc} >$ 5.2 GeV/$c^2$ and 
$|\Delta E| <$ 0.2 GeV.
For the $\phi\phi K^\pm$ mode, we simultaneously obtain the yield and the partial rate asymmetry $\mathcal{A}_{CP}$ using the likelihood, defined as:  

\begin{eqnarray}
\mathcal{L} = {\rm exp}[-(N_S + N_B)] \prod_i^N (\sum_j \frac{1}{2}[1-q_i\cdot {\mathcal A}^j_{CP}] N_j P_i^j(M_{\rm bc},\Delta E)),
\label{eq: ML} 
\end{eqnarray}
where $i$ is the identifier of the $i$-th event, $j$ indicates signal or background, $P(M_{\rm bc},\Delta E)$ is the two-dimensional PDF of $M_{\rm bc}$ and $\Delta E$, and $q$ indicates the $B$ meson flavor, $+1$ for $B^+$ and $-1$ for $B^-$, respectively.
For neutral $B$ events, $\frac{1}{2}[1-q_i\cdot {\mathcal A}_{CP}]$ in Eq. (\ref{eq: ML}) is replaced by 1.
The $M_{\rm bc}$ PDFs are modeled by a Gaussian function for signals and an ARGUS function~\cite{argus} for the continuum, while a Gaussian is used to describe the signal $\Delta E$ and a $2^{nd}$ order Chebyshev polynomial is used for the background $\Delta E$.
The parameters of the PDFs are determined using high-statistics MC samples and sideband data for signal and background shapes.
The signal PDFs are calibrated by comparing the $M_{\rm bc}$ and $\Delta E$ distributions of the $B^{+} \to \overline{D}{}^{0}(K^+\pi^-\pi^-\pi^+) \pi^{+}$ events with the MC expectation.\par

\begin{figure}[htb]
\includegraphics[scale=0.6]{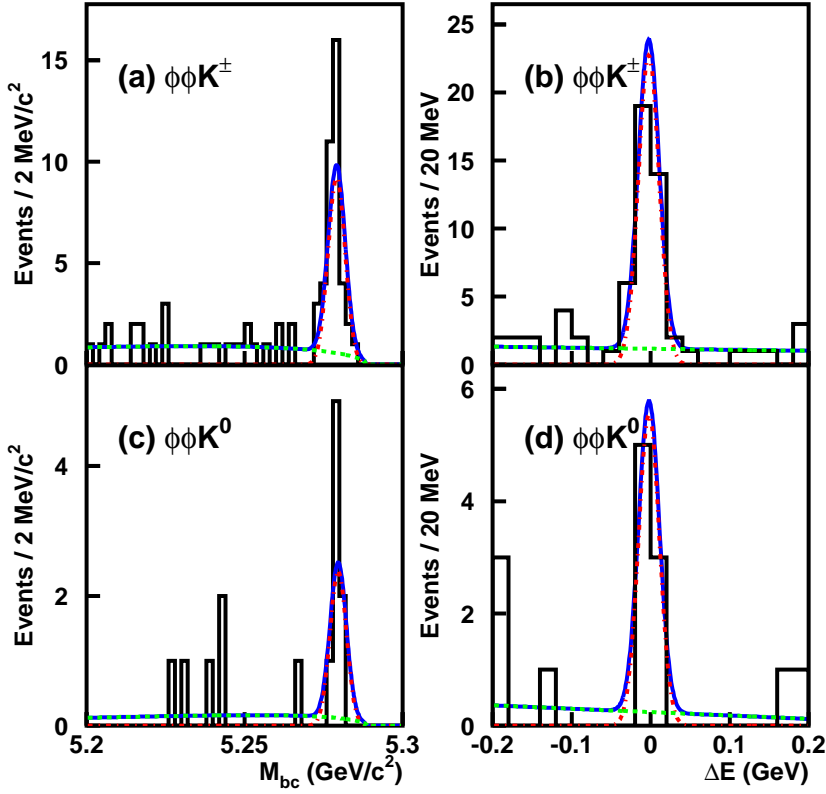}
\caption{Distributions of $M_{\rm bc}$ and $\Delta E$ for the decay modes $B^{\pm} \to \phi \phi K^{\pm}$ (a,b) and $B^{0} \to \phi \phi K^{0}$ (c,d), with $\phi \phi$ invariant mass less then 2.85 GeV/$c^2$.
The open histograms represent the data, the solid blue curves show the result of the fit,  the dash-dotted red lines represent the signal contributions and the dashed green curves show the continuum background contributions.}
\label{fig: phiphik_fig1}
\end{figure}

\begin{figure}[htb]
\includegraphics[scale=0.6]{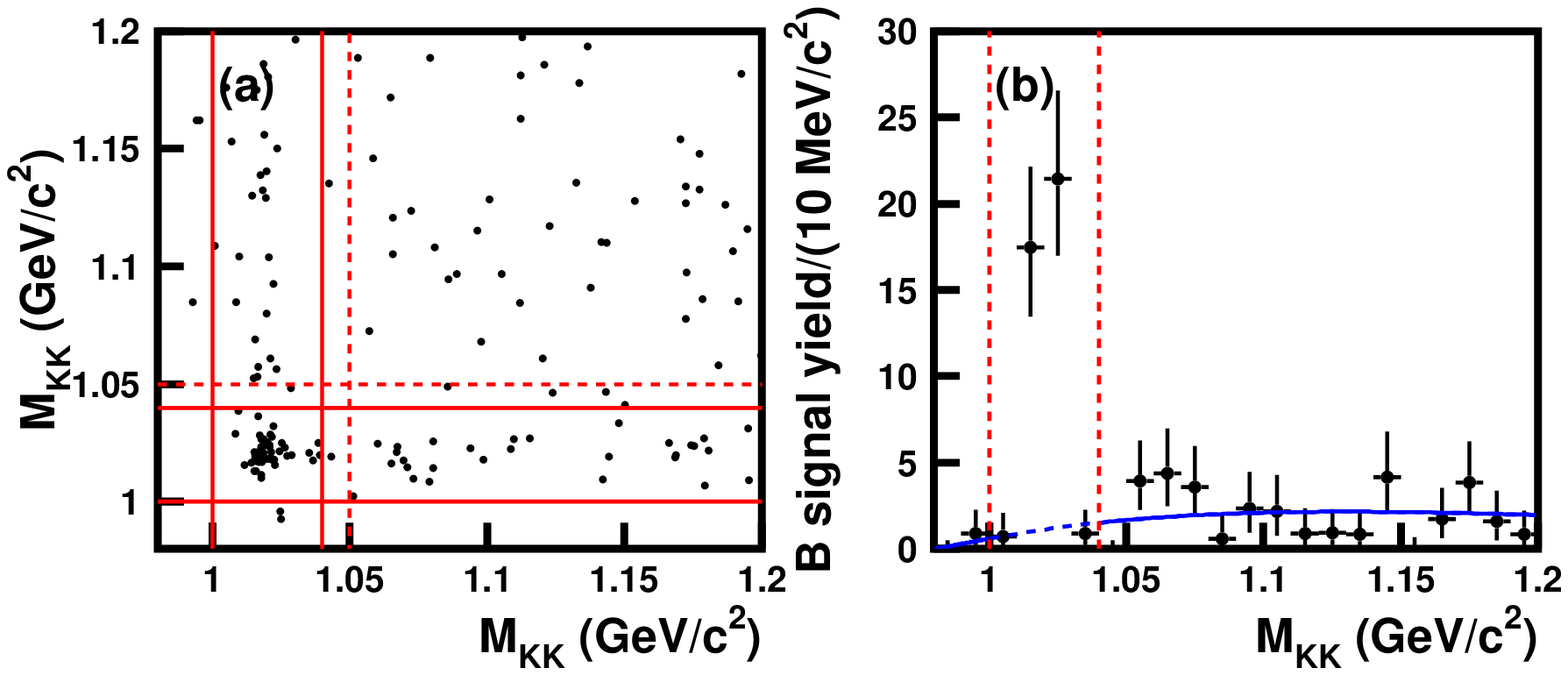}
\caption{(a) The distribution of $M_{K^+K^-}$ vs. $M_{K^+K^-}$ for the $K^+K^-K^+K^-K^\pm$ candidates in the $M_{\rm bc}-\Delta E$ signal box with $M_{K^+K^-} <$ 1.2 GeV/$c^2$.
The two $K^+K^-$ bands indicate the $\phi$ mass region (1.0 GeV/$c^2 < M_{K^+K^-} <$ 1.04 GeV/$c^2$).
The horizontal and vertical dashed lines are located at $M_{K^+K^-}$ = 1.05 GeV/$c^2$.
The rectangle on the upper right is the $\phi \phi$ sideband region; events in this region are used to estimate the non-resonant $B\to 5K$ contribution.
(b) $B$ signal yields as a function of the $M_{K^+K^-}$ after requiring the other $K^+K^-$ pair has a mass in the $\phi$ mass region.
The threshold function is used to be the fit curve and events with 1.0 GeV/$c^2 < M_{K^+K^-} <$ 1.04 GeV/$c^2$ are excluded when the fit is performed.}
\label{fig: phiphik_fig2}
\end{figure}

We search for charmless $B\to \phi\phi K$ decays by requiring the $\phi \phi$ invariant mass ($M_{\phi \phi}$) to be less than 2.85  GeV/$c^2$, the region below the charm threshold.
Candidate $\phi$ mesons are identified by requiring the invariant masses of $K^+K^-$ pairs ($M_{K^+K^-}$) to be in the range 1.0 GeV/$c^2$ to 1.04 GeV/$c^2$ ($\pm 4.6\sigma$).
Figure \ref{fig: phiphik_fig1} shows the $M_{\rm bc}$ and $\Delta E$ projections with the fit curves superimposed. 
Clear signals appear in both $B^\pm$ and $B^0$ modes with signal yields of $37.0^{+6.7}_{-6.0}$ and $7.8^{+3.2}_{-2.5}$, respectively.
Although $K^+K^-$ candidates are required to lie in the $\phi$ mass region, non-$\phi$ backgrounds may also contribute.
Figure \ref{fig: phiphik_fig2}(a) shows the $M_{K^+K^-}$ vs. $M_{K^+K^-}$ distributions for $(K^+K^-K^+K^-)K^\pm$ candidates in the signal region, where the two $K^+K^-$ pairs are required to have invariant masses less than 1.2 GeV/$c^2$.
Events in the two $\phi$ bands are used to estimate the $B^\pm\to\phi K^+K^-K^\pm$ contribution.
Figure \ref{fig: phiphik_fig2}(b) shows $B$ signal yields~\cite{B signal yields} as a function of  the $K^+K^-$ invariant mass after requiring the other $K^+ K^-$ pair to have a mass in the $\phi$ mass region.
The $B$ signal yields are fitted with a threshold function in the region 0.98 GeV/$c^2 < M_{K^+K^-} <$ 1.2 GeV/$c^2$, excluding $\phi$ mass region (1.0 GeV/$c^2 < M_{K^+K^-} <$ 1.04 GeV/$c^2$).
The amount of the  non-$\phi$ contribution is estimated by interpolating the $B$ yields in the $\phi$ sideband region to the $\phi$ mass region, which is $4.4^{+0.8}_{-0.7}$ events.
Since events in the two $\phi$ bands contain both true $\phi$ mesons and non-resonant $K^+K^-$ pairs, the area underneath the $\phi$ mass region in Fig. \ref{fig: phiphik_fig2}(b) includes the $\phi K^+K^-K^\pm$ contribution but counts the non-resonant $5 K$ component twice.
Therefore, we estimate the non-resonant $B\to 5 K$ contribution using the $B$ signal yield in the upper right corner of the dashed region in Fig. \ref{fig: phiphik_fig2}(a).
We assume a phase-space distribution in 4-kaon mass.
We obtain $1.3\pm 0.4$ non-resonant events in the $\phi\phi K^\pm$ sample.
After subtracting this contribution of $1.3\pm 0.4$ events, the non-$\phi\phi K$ fraction  is calculated to be ($7\pm 4$)\%.
The same procedure is applied to the $\phi\phi K^0$ sample; here we obtain a fraction of ($7 \pm 9$)\%.\par

Table \ref{table: phiphik results} summarizes the $\phi \phi K$ results after subtracting the non-$\phi \phi K$ contribution.
Signal efficiencies are obtained by generating $\phi \phi K$ MC events, where the same $M_{\phi \phi} <$ 2.85 GeV/$c^2$ requirement is applied.
Systematic uncertainties in the fit are obtained by performing fits in which the signal peak positions and resolutions of the signal PDFs are successively varied by $\pm 1 \sigma$.
The quadratic sum of each deviation from the central value of the fit gives the systematic uncertainty of the fit.
For each systematic check, the statistical significance is taken as $\sqrt{-2\ln(\mathcal{L}_{\rm feeddown}/\mathcal{L}_{\rm max})}$, where $\mathcal{L}_{\rm feeddown}$ and $\mathcal{L}_{\rm max}$ are the likelihoods at the expected non-$\phi\phi K$ yields and the best fit, respectively.
We regard the smallest value as our significance including the systematic uncertainty.
The number of $B^+B^-$ and $B^0\overline{B}{}^0$ pairs are assumed to be equal.\par

\begin{table}[htb]
\caption{Signal yields, efficiencies including secondary branching fractions, significances, branching fractions for $B \to \phi \phi K$ and related charmonium decays.}
\label{table: phiphik results}
\begin{footnotesize}
\begin{tabular}
{@{\hspace{3mm}}l@{\hspace{3mm}}@{\hspace{3mm}}c@{\hspace{3mm}}@{\hspace{3mm}}c@{\hspace{3mm}}@{\hspace{3mm}}c@{\hspace{3mm}}@{\hspace{3mm}}c@{\hspace{3mm}}@{\hspace{3mm}}c@{\hspace{3mm}}}
\hline \hline
Mode & Yields & efficiencies(\%) & $\Sigma$ & $\mathcal{B} (10^{-6})$ \\
\hline
$B^{\pm} \to \phi \phi K^{\pm}$  ($M_{\phi \phi} <$ 2.85 GeV/$c^2$) & $34.2^{+6.4}_{-5.8}$ & 2.41 & 9.5 & $3.2^{+0.6}_{-0.5} \pm 0.3$ \\
$B^{0} \to \phi \phi K^{0}$ ($M_{\phi \phi} <$ 2.85 GeV/$c^2$) & $7.3^{+3.0}_{-2.4}$ & 0.69 & 4.7 & $2.3^{+1.0}_{-0.7} \pm 0.2$ \\
$B^\pm \to \eta_c K^\pm$, $\eta_c \to \phi \phi$ & $29.7^{+6.8}_{-5.5}$ & 2.72 & 7.2 & $2.4^{+0.6}_{-0.5}\pm 0.2$ \\
$B^\pm \to \eta_c K^\pm$, $\eta_c \to \phi K^+K^-$ & $76.8^{+13.6}_{-12.4}$ & 4.85 & 9.4 & $3.5\pm 0.6 \pm 0.3$ \\
$B^\pm \to \eta_c K^\pm$, $\eta_c \to 2(K^+K^-)$ & $104.6^{+20.2}_{-17.3}$ & 9.93 & 10.2 & $2.4^{+0.5}_{-0.4}\pm 0.2$ \\
$B^\pm \to J/\psi K^\pm$, $J/\psi \to \phi K^+K^-$ & $25.5^{+7.0}_{-6.0}$ & 4.67 & 8.5 & $1.2\pm 0.3\pm 0.1$ \\
$B^\pm \to J/\psi K^\pm$, $J/\psi \to 2(K^+K^-)$ & $41.0^{+7.3}_{-6.6}$ & 9.41 & 9.7 & $0.97^{+0.17}_{-0.16}\pm 0.1$ \\ 
\hline \hline
\end{tabular}
\end{footnotesize}
\end{table}

The performance of the $\mathcal{R}$ requirement is studied by checking the data-MC efficiency ratio using the $B^{+} \to \overline{D}{}^{0}(\to K^{+} \pi^{-} \pi^{-} \pi ^{+})\pi^{+}$ sample.
The obtained error is 2.7-2.8\%.
The systematic errors on the charged track reconstruction are estimated to be around $1$\% per track using  partially reconstructed $D^*$ events.
Therefore, the tracking systematic error is 5\% (5 tracks) for the $\phi\phi K^\pm$ mode and 4\% for the $\phi\phi K^0$ mode.
The kaon identification efficiency is studied using samples of inclusive $D^{*+}\to D^0\pi^+, D^0\to K^-\pi^+$ decays.          
The $K_S^0$ reconstruction is verified by comparing the  ratio of $D^+\to K_S^0\pi^+$ and $D^+\to K^-\pi^+\pi^+$ yields.
The resulting $K_S^0$ detection systematic error is 4.9\%.
The uncertainty in the number of $B\overline{B}$ events is 1\%. 
The final systematic error is obtained  by summing  all correlated errors linearly and then quadratically summing the uncorrelated errors.\par

After subtracting the non-$\phi\phi K$ contribution, the branching fractions for charmless $B \to \phi \phi K$ decays are $\mathcal{B}(B^{\pm} \to \phi \phi K^{\pm}) = (3.2^{+0.6}_{-0.5} \pm 0.3) \times 10^{-6}$ with 9.5$\sigma$ significance and $\mathcal{B}(B^{0} \to \phi \phi K^{0}) = (2.3^{+1.0}_{-0.7} \pm 0.2) \times 10^{-6}$ with 4.7$\sigma$ significance.
The measured charge asymmetry for $B^{\pm} \to \phi \phi K^{\pm}$ decay is $0.01^{+0.19}_{-0.16} \pm 0.02$.
The first error is statistical and the second is systematical.\par

\begin{figure}[htb]
\includegraphics[scale=0.6]{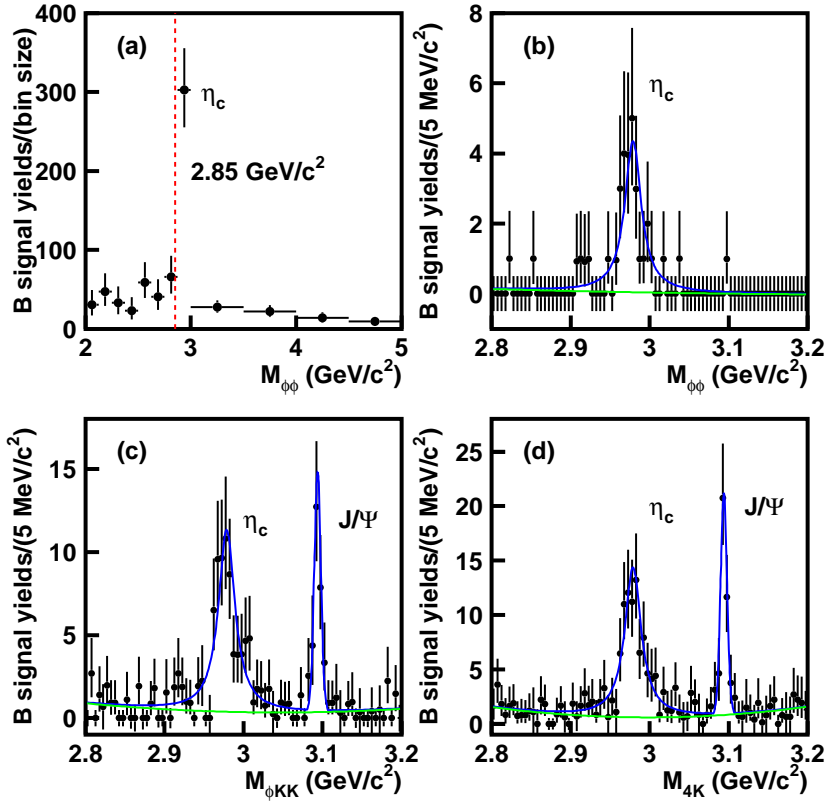}
\caption{$B$ signal yields as a function of (a,b) $M_{\phi \phi}$, (c) $M_{\phi K^{+}K^{-}}$ and (d) $M_{4K}$.
In (a) we use different bin sizes for $M_{\phi \phi}$ less than 3 GeV/$c^2$ and greater than 3 GeV/$c^2$.
The subset with $M_{\phi \phi}$ from 2.8 GeV/$c^2$ to 3.2 GeV/$c^2$ is shown in (b).}
\label{fig: phiphik_fig3}
\end{figure}

It is of interest to search for possible $\phi \phi$ resonances above charm threshold.
Figure \ref{fig: phiphik_fig3}(a) shows the $B$ signal yields divided by the bin size as a function of $M_{\phi \phi}$ after releasing the $M_{\phi \phi} <$ 2.85 GeV/$c^2$ requirement.
There is no specific enhancement in high $\phi \phi$ mass region except for the $\eta_{c}$ peak near 3 GeV/$c^2$.
Reference \cite{Hazumi} suggests a large $CP$ asymmetry, that arises from the interference between $B^\pm \to \phi \phi K^\pm$ and $B^\pm \to \eta_c(\to \phi \phi) K^\pm$ decays.
The events with $\phi \phi$ invariant mass within $\pm 40$ MeV/$c^2$ of the nominal $\eta_c$ mass are selected to investigate this asymmetry.
The measured $CP$ asymmetry is $0.15^{+0.16}_{-0.17} \pm 0.02$, which is consistent with no asymmetry.\par

We study possible charmonium states by checking $B$ yields with $M_{4K}$ between 2.8 GeV/$c^2$ and 3.2 GeV/$c^2$.
Since $\eta_c$ and $J/\psi$ mesons may decay to $\phi K^+K^-$ and  $2(K^+K^-)$ pairs, a mass scan is performed with and without the requirement that the $K^+K^-$ pair lie in the $\phi$ mass region.
As shown in Fig. \ref{fig: phiphik_fig3}, clear $\eta_c$ and $J/\psi$ resonances are visible in the $\phi K^+K^-$ and $4K$ samples while only an $\eta_c$ peak appears in the $\phi \phi$ mode.\par

We obtain the signal yields for $B^\pm\to \eta_c K^\pm$ and $B^\pm\to J/\psi K^\pm$ by performing binned histogram fits to Figs. \ref{fig: phiphik_fig3}(b), \ref{fig: phiphik_fig3}(c) and \ref{fig: phiphik_fig3}(d).
The $J/\psi$ signal PDF is modeled with a Gaussian function while the $\eta_c$ PDF is described by a Breit-Wigner function convolved with a Gaussian resolution function, which has the same Gaussian width as the $J/\psi$ PDF.
Since sizable signals are observed in the $4K$ mode, the parameters are determined using the $4K$ sample and the same signal PDFs are then applied to the $\phi K^+K^-$ and $\phi \phi$ samples.
The obtained Gaussian width is measured to be $4.0^{+1.0}_{-0.8}$ MeV/$c^2$ by performing a fit with a $2^{nd}$ order Chebyshev polynomial as the non-resonant PDF after excluding events in the $\eta_c$ mass region (2.94 GeV/$c^2 < M_{4K} < 3.02$  GeV/$c^2$).
A fit is performed to the full range with the Gaussian width fixed.
The signal yields are summarized in Table \ref{table: phiphik results}.
The peak positions obtained for the $\eta_c$ and $J/\psi$ are $2.979\pm 0.002$ GeV/$c^2$ and $3.094\pm 0.001$ GeV/$c^2$, respectively, consistent with the nominal $\eta_c$ and $J/\psi$ masses.
The $\eta_c$ Breit-Wigner width is measured to be $25.2^{+7.7}_{-6.0} \pm 0.3$ MeV/$c^2$, where the central value is consistent with the world average and the second error is due to the uncertainty in the mass resolution.\par

For the $\phi K^+K^-$ and $\phi \phi$ modes, the non-$\phi$ contribution is determined from the $B$ signal yields for events with one $K^+K^-$ pair in the $\phi$ sideband region (1.05 GeV/$c^2 <M_{K^+K^-} <$ 1.09 GeV/$c^2$) and the $4K$ and $\phi K^+K^-$ masses are in the charmonium resonance region, respectively.
We find $3.0^{+2.0}_{-1.4}$ events in the $\eta_c \to \phi \phi$ mode, $6.4^{+5.4}_{-4.4}$ events in the $\eta_c \to \phi K^+ K^-$ mode, and $3.4^{+3.6}_{-2.6}$ in the $J/\psi \to \phi K^+ K^-$ mode.
After subtracting the feed-down yields, we obtain the results listed in Table \ref{table: phiphik results}.\par

Signal efficiencies are determined using signal MC and the detection systematic uncertainties are similar to what was described in the charmless 
$\phi\phi K$ part.
Fit systematic uncertainties are estimated by successively varying the peak positions and resolutions of the $M_{\rm bc}-\Delta E$ signal PDFs as well as the convolution Gaussian width in the fit.
The quadratic sum of each deviation gives the fit systematic errors.
Since the sub-decay branching fractions of $\eta_c$ and $J/\psi$ mesons to $4K, \phi KK$ and $\phi\phi$ final states are not precisely known, we provide the product of branching fractions for various decays in Table \ref{table: phiphik results}.
Using the known branching fractions of $\mathcal{B}(B^{\pm} \to \eta_{c} K^{\pm}) = (9.1 \pm 1.3) \times 10^{-4}$ and $\mathcal{B}(B^{\pm} \to J/\psi K^{\pm}) = (1.008 \pm 0.035) \times 10^{-3}$ \cite{PDG}, the subdecay branching fractions are calculated and listed in Table \ref{table: subdecay BF}.\par

\begin{table}[htb]
\caption{The measured branching fractions of secondary charmonium decays and the world averages \cite{PDG}.}
\label{table: subdecay BF}
\begin{tabular}
{@{\hspace{0.5cm}}l@{\hspace{0.5cm}}@{\hspace{0.5cm}}c@{\hspace{0.5cm}}@{\hspace{0.5cm}}c@{\hspace{0.5cm}}}
\hline \hline
Decay mode & $\mathcal{B}$ (measured) & $\mathcal{B}$ (PDG 2006 value) \\
\hline
$\eta_{c} \to \phi \phi$ & $(2.7^{+0.6}_{-0.5} \pm 0.4) \times 10^{-3}$ & $(2.7 \pm 0.9) \times 10^{-3}$ \\
$\eta_{c} \to \phi K^{+}K^{-}$ & $(3.9^{+0.7}_{-0.6} \pm 0.6) \times 10^{-3}$ & $(2.9 \pm 1.4) \times 10^{-3}$ \\
$\eta_{c} \to 2(K^{+}K^{-})$ & $(2.6^{+0.5}_{-0.4} \pm 0.4) \times 10^{-3}$ & $(1.5 \pm 0.7) \times 10^{-3}$ \\
$J/\psi \to \phi K^{+}K^{-}$ & $(1.2 \pm 0.3 \pm 0.1) \times 10^{-3}$ & $(1.83 \pm 0.24) \times 10^{-3}$ \\
$J/\psi \to 2(K^{+}K^{-})$ & $(9.7^{+1.7}_{-1.6} \pm 1.0) \times 10^{-4}$ & $(7.8 \pm 1.4) \times 10^{-4}$ \\
\hline \hline
\end{tabular}
\end{table}

In summary, we have observed the charmless decay $B^{\pm} \to \phi \phi K^{\pm}$ and evidence of $B^{0} \to \phi \phi K^{0}$.
We also report the $CP$ asymmetry of the charged decay and measurements of other closely related charmonium decays.
The results are consistent with the previous measurements, but have considerably improved precision due to the increase in statistics.


We thank the KEKB group for the excellent operation of the
accelerator, the KEK cryogenics group for the efficient
operation of the solenoid, and the KEK computer group and
the National Institute of Informatics for valuable computing
and Super-SINET network support. We acknowledge support from
the Ministry of Education, Culture, Sports, Science, and
Technology of Japan and the Japan Society for the Promotion
of Science; the Australian Research Council and the
Australian Department of Education, Science and Training;
the National Science Foundation of China and the Knowledge 
Innovation Program of the Chinese Academy of Sciences under 
contract No.~10575109 and IHEP-U-503; the Department of Science 
and Technology of India; the BK21 program of the Ministry of Education of
Korea, and the CHEP SRC program and Basic Research program 
(grant No. R01-2005-000-10089-0) of the Korea Science and
Engineering Foundation; the Polish State Committee for
Scientific Research under contract No.~2P03B 01324; the
Ministry of Science and Technology of the Russian
Federation; the Slovenian Research Agency;  
the Swiss National Science Foundation; the National Science Council and
the Ministry of Education of Taiwan; and the U.S.\
Department of Energy.


%


\begin{thebibliography}{99}
\bibitem{Huang} 
H.C.~Huang {\it et al.} (Belle Collaboration),
Phys. Rev. Lett. {\bf 91}, 241802 (2003).

\bibitem{Hazumi}
M.~Hazumi,
Phys. Lett {\bf B 583}, 285 (2004).

\bibitem{KEKB}
S.~Kurokawa and E.~Kikutani, Nucl. Instr. and. Meth. A499, 1 (2003),
and other papers included in this volume.

\bibitem{Belle}
A.~Abashian {\it et al.} (Belle Collaboration),
Nucl. Instr. and Meth. A {\bf 479}, 117 (2002).

\bibitem{Ushiroda} Y. Ushiroda,
Nucl. Instr. and Meth. A {\bf 511} 6 (2003).

\bibitem{fisher}
R. A. Fisher, Ann. Eugenics {\bf 7}, 179 (1936).

\bibitem{SFW}
The Fox-Wolfram moments were introduced in G.~C.~Fox and S.~Wolfram, Phys. Rev. Lett. {\bf 41}, 1581 (1978).
The Fisher discriminant used by Belle, based on modified Fox-Wolfram moments (SFW), is described in K.~Abe {\it et al.} (Belle Collaboration.), Phys. Rev. Lett. {\bf 87}, 101801 (2001) and K.~Abe {\it et al.} (Belle Collabboration.), Phys. Lett. {\bf B 511}, 151 (2001).

\bibitem{spher} R. Ammar {\it et al.} (CLEO Collaboration),
 Phys. Rev. Lett. {\bf 71}, 674 (1993).

\bibitem{TaggingNIM}
H. Kakuno {\it et al.}, Nucl. Instr. and Meth.A {\bf 533} 516 (2004).

\bibitem{argus}
H. Albrecht {\it et al.} (ARGUS Collaboration), Phys. Lett. B {\bf 229}, 304 (1989).

\bibitem{B signal yields}
Hereafter in this paper, the $B$ signal yield is obtained from 2D $M_{\rm bc}-\Delta E$ fits to the events in each bin of the plot.

\bibitem{PDG}
W.-M. Yao {\it et al.} (Particle Data Group), 
Journal of Physics G 33, 1 (2006)

\end{thebibliography}
\end{document}